\documentstyle[editedvolume]{jd} 
\begin{opening}
\title{DYNAMICAL SIMULATIONS: METHODS AND COMPARISONS}

\author{D.C. HEGGIE}
\institute{Department of Mathematics and Statistics, University of Edinburgh,\\
	   King's Buildings, Edinburgh EH9 3JZ, U.K.}

\author{Mirek Giersz}
\institute{N. Copernicus Astronomical Center, Polish Academy of Sciences,\\
           ul. Bartycka 18, 00-716 Warsaw, Poland}

\author{Rainer Spurzem}
\institute{Astronomisches Rechen-Institut, Moenchhofstrasse 12-14,\\
           D-69120 Heidelberg, Germany}

\author{Koji Takahashi}
\institute{Department of Earth and Space Science, Osaka University\\
           Toyonaka, Osaka 560, Japan}

\end{opening}

\runningtitle{DYNAMICAL SIMULATIONS: METHODS AND COMPARISONS}

\begin{document}


\section{A Comparative Assessment of Dynamical Models by D.C. Heggie}

\subsection{Introduction}

In the dynamical study of globular star clusters, five types of
dynamical models are in current use.  The following list includes recent
highlights, some of which are developed in these proceedings by other
authors.

\begin{enumerate}
\item \emph{Static models}: Besides the traditional King models and their
variants, \emph{non-parametric} models are of increasing importance (cf.
Meylan, these proceedings).  Dynamical evolution has no role in static
models, however.

\item \emph{Scaling models}, in which a cluster is assumed to evolve
along a simple sequence like the King sequence, which can be
characterised in terms of a length and mass scale (and perhaps other
parameters, such as the concentration.)  These have not been used much
since the work of Chernoff \& Shapiro (1987).

\item \emph{Gas models}:  Though little used for the modelling of
observations, these models have played a major role in theory, as the
phenomena of core collapse and gravothermal oscillations were first
developed in this context;  cf. Spurzem, this paper, Sect. 4.

\item \emph{Fokker-Planck models}  remain the principal tool for
studying the dynamical evolution of both individual clusters (e.g. Sosin
\& King 1997) and the cluster system as a whole (e.g. Gnedin \& Ostriker
1997, Murali \& Weinberg 1997).  Recent exciting developments are
described in Einsel \& Spurzem (1997) and in contributions to this
paper by Takahashi (Sect. 2) and by Giersz (Sect. 3).

\item \emph{$N$-body models} are increasingly directed towards study of
globular clusters (with suitable scaling), thanks to spectacular
hardware developments.   Along with applications and future
developments, these are described in contributions to these proceedings
by Makino, Taiji, Vesperini, Tout and Portegies Zwart.
\end{enumerate}

\subsection{Intercomparison of Dynamical Models}

All the above kinds of models depend on simplifying assumptions, and
their reliability may be investigated by studying the same problems with
different models.  For example, observations of mass segregation in star
clusters are usually interpreted in terms of static, multi-mass King
models, and so a comparison of such models with evolved multi-mass
Fokker-Planck models is a guide to their reliability.  Though some
comments on this are to be found in the literature (e.g. Chernoff \&
Weinberg 1990, Sosin \& King 1997), a systematic study is overdue.

Comparisons between the three main evolutionary models (gas,
Fokker-Planck and $N$-body) are more plentiful (e.g. Aarseth, H\'enon \&
Wielen 1974, Bettwieser \& Inagaki 1985, Giersz \& Spurzem 1994, Spurzem
\& Takahashi 1995, Spurzem \& Aarseth 1996, Theuns 1996). 
Nevertheless most of these comparisons concern single-mass systems, and
almost all deal with isolated clusters.  

Because of the interest in a more realistic comparison, in recent months
a number of groups have collaborated by working with a variety of codes
on a single, somewhat more elaborate model.  Initially the model is a
non-rotating King model (with $W_0 = 3$), with a total mass of
$6\times10^4M_\odot$ consisting entirely of single point masses with a
Salpeter mass function in the range $0.1<m/M_\odot<1.5$ and no initial
mass segregation.  The cluster is in a circular galactic orbit of radius
$\simeq 4.2$kpc with a tidal radius of $30$pc.  Heating is by three-body
binaries (which are formed around and after core bounce).  No form of
stellar evolution is included.

This prescription is a compromise between realism and the limitations of
existing codes.  In the event, three $N$-body codes, six Fokker-Planck
codes and one gas code were able to produce data, thanks to the efforts
of S.J. Aarseth, G.A. Drukier, C.R.W. Einsel, K. Engle, T. Fukushige, M.
Giersz, D.C. Heggie, P. Hut, H.-M. Lee, J. Makino, S.L.W. McMillan,
S.F. Portegies Zwart, G. Quinlan, R. Spurzem and K. Takahashi.  Some
detailed results can be seen at
http://www.maths.ed.ac.uk/people/douglas/experiment.html, and an
extended account of the collaboration is in preparation.  Here we
summarise some points of interest.

\begin{figure}[htb]
\vspace{8cm} 
\includegraphics{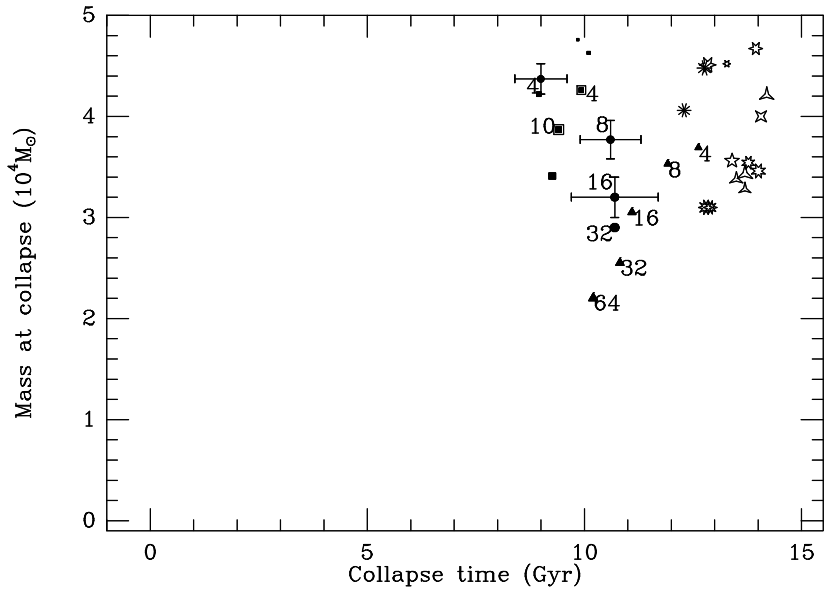} 
\caption{Mass within the tidal radius at
the time of core bounce, plotted against the time at which core bounce
occurred.  The symbols have the following meanings:  star -- anisotropic
gas model; open symbols -- Fokker-Planck models;  filled symbols --
$N$-body models.  Symbols of the same shape correspond to results from
the same group or individual.  Larger symbols denote ``better'' models,
corresponding to larger $N$ (for $N$-body models) or larger
numbers of mass groups and/or smaller timesteps 
(for continuum models).  In some cases different
symbols correspond to different treatments of the tidal boundary
condition; for the $N$-body models with square symbols, the models with
a circumscribed rectangle use the same boundary conditions as the
$N$-body models denoted by triangles and disks.  Most of the latter are
obtained from several runs with the same $N$, and the error bars on
those cases denote $1\sigma$.  All models have been reduced to a common
choice of the Coulomb logarithm ($\ln 0.01N$) in the expression for the
relaxation time.  The value of $N$ is given (in units of $1024$) beside 
the main $N$-body models.The open, three-sided symbols give four
results for Takahashi's Fokker-Planck models; the most discrepant (with
the highest mass) is an anisotropic model with apocentric escape
criterion (cf. his contribution to these proceedings).}
\end{figure}

All models in the present study indicate core collapse at around 10Gyr
and those that reached dissolution did so by around 20Gyr.  Fig.1 shows
data on the time of core collapse and the total mass at that time (which
relates to the rate of mass loss and hence the dissolution time).  All
the continuum models produce a fairly consistent collapse
time, as do the largest $N$-body models.  The mean collapse time of the
$N$-body models is earlier than that for the continuum models, but most
of the discrepancy could be removed by changing the Coulomb logarithm to
about $\ln 0.06N$.

The mass at core bounce is independent of the choice of Coulomb
logarithm.  All kinds of models show considerable variation, especially
the $N$-body models.  What is especially interesting about the latter is
that the results vary systematically with $N$, in the sense that larger
models lose mass more quickly, and that the trend is
similar in all three sets of models (which were obtained by independent
groups using independently written codes).  Also, there is little
indication that the results are converging with increasing $N$.
The actual number of stars in the cluster which is being simulated is
approximately $2.5\times10^5$, and a rough extrapolation of the $N$-body
results with $N$ suggests that the mass at core bounce could be as small
as $1.5\times10^4M_\odot$.  If so, it follows that existing $N$-body models
and all continuum models are in error (for this model) by a factor of
about two, in the sense that they lose mass too slowly.

One of the most interesting discoveries from this investigation is the
aforementioned trend of the $N$-body models with $N$. No such trend
should occur if the evolution scales with the relaxation time, as is
assumed by the continuum models.  It may be that the Coulomb logarithm
must be chosen with greater care: it may involve the number of stars in
the core, or the interaction between different pairs of masses may
require different values.  It may be that there are other processes
(related to escape or the anisotropic tidal field) whose time scales do
not scale with the relaxation time.  Uncovering the reason for the trend
with $N$ will require further research.  In the meantime, the time scale
for cluster evolution obtained from continuum and $N$-body models should
be treated with caution.
\bigskip

This work was
supported in part by the UK Particle Physics and Astronomy Research
Council under grant GR/J79461;  this grant funded GRAPE hardware which
was kindly supplied by the group of Prof. D. Sugimoto, University of
Tokyo.

\bigskip

\section{Anisotropic Fokker-Planck Models of 
       Globular Cluster Evolution by K. Takahashi}

\subsection{Introduction}

The dynamical evolution of globular clusters driven by two-body
relaxation
was investigated by numerical integration of the two-dimensional
Fokker-Planck (FP) equation in energy--angular momentum space
(Cohn 1979, 1980).
The two-dimensional FP models allow anisotropy of the velocity
distribution of stars.
In this paper, we report
the results of our FP simulations of the evolution of globular clusters,
and discuss in particular the development of velocity anisotropy
and its effects on the cluster evolution.

If spherical symmetry and dynamical equilibrium of a cluster are
assumed, then the distribution function ($f$) is a function of the
energy per unit mass ($E$), the modulus of the angular momentum per unit
mass ($J$), and time ($t$).  If isotropy of the velocity distribution is
assumed furthermore, then $f$ is a function of only the energy and time.
Anisotropic FP models are, of course, more realistic models of globular
clusters than isotropic models.

Numerical integration of the FP equation was performed by using
a method described in Takahashi (1995, 1996).
We considered the mass spectrum of stars (Takahashi 1997)
and the effect of tidal truncation (Takahashi et al. 1997)
in order to investigate the realistic evolution of globular clusters.

\subsection{Results}

In isolated single-mass clusters, the halo rapidly becomes dominated by
radial orbits.
The ratio of the radial velocity dispersion to the tangential one
increases monotonically as the radius increases.
Also in isolated multi-mass clusters,
the radial anisotropy develops in the halo.
However, the radial profiles of the velocity anisotropy
are significantly different between different masses in some cases.
For example, a strong tangential anisotropy can develop around
the half-mass radius for massive components in a cluster with a steep
mass spectrum.
In tidally truncated clusters, although the radial
anisotropy develops in the halo during the pre-collapse evolution,
the anisotropy becomes highly depressed during the post-collapse
evolution
due to rapid loss of radial orbits.
When the tidal field is weak, 
the cluster loses mass faster in the anisotropic model
than in the isotropic model.
However,
the difference in mass loss rate between the two models
becomes smaller as the strength of the tidal field increases.
This is because there is not enough room for the radial anisotropy
to develop when the tidal field is strong.

\subsection{Discussion}

We adopted two different tidal-cutoff conditions in $(E,J)$ space:
``apocenter condition" and ``energy condition" (Takahashi et al. 1997).
The apocenter condition removes stars whose apocenter radii are greater
than the tidal radius.  The energy condition removes stars whose
energies are greater than the tidal energy (or the potential energy at
the tidal radius).  Intuitively the apocenter condition seems to be a
more realistic cutoff condition.  However, a comparison with $N$-body
simulations (cf. Heggie's talk in this session) does not necessarily
support this idea.  Further investigation is needed on this problem.

\bigskip
This work was supported in part by the Grant-in-Aid for Encouragement of Young
Scientists by the Ministry of Education, Science, Sports and Culture of
Japan (No. 1338).

\bigskip

\section{Monte-Carlo Simulations. First Results by M. Giersz}

\subsection{Introduction}

A revision of Stod\'o\l kiewicz's Monte--Carlo code (Stod\'o\l kiewicz
1982, 1985, 1986) was used to simulate the evolution of star
clusters. The Monte--Carlo method can be regarded as a statistical way
of solving the Fokker--Planck equation. The great advantages of this
method, beside of its simplicity and speed, are connected with the
inclusion of anisotropy and with the fact that added realism does not
slow it down. The Monte--Carlo method can practically cope as easily as
the $N$--body method with internal degrees of freedom of single and
binary stars and external environment, with one exception, a stellar
system must be spherically symmetric. The new method treats each {\it
superstar} as a single star and follows the evolution and motion of all
individual stellar objects. This enables, for example, proper
description of the densest parts of the system and mass segregation of
binaries.

\begin{figure}[htb]
\vspace{8cm} 
\includegraphics{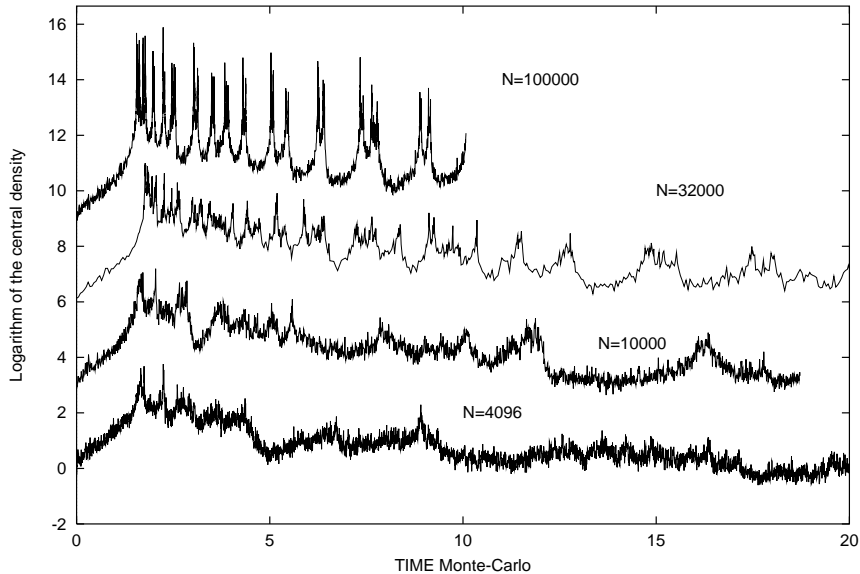} 
\caption{Evolution of the central density for $N = 4096$, $10000$, 
$32000$ and $100000$. Data are shifted in the logarithm by $3$, $6$ and 
$9$ for $N = 10000$, $32000$ and $100000$, respectively.} 
\end{figure}

\subsection{First results}

The first calculations (Giersz 1997) for equal--mass $N$--body systems
with three--body energy generation according to Spitzer's formulae show
good agreement with direct $N$--body calculations for $N=2000$, $4096$
and $10000$ particles. The density, velocity, mass distributions, energy
generation, number of binaries etc. follow the $N$--body results. The
only difference is that there is no levelling off of the anisotropy
level for advanced post--collapse evolution of Monte--Carlo models as is
seen in $N$--body simulations for $N \leq 2000$.  For simulations with
$N \geq 10000$ gravothermal oscillations are clearly visible (see
Fig.2). This is the first unambiguous detection of gravothermal
oscillations in Monte--Carlo simulations. Moreover, this is a first
unambiguous detection of gravothermal oscillations for a stochastic
$N$--body system with $N$ as large as $100000$. It should be noted that
for oscillations observed in Monte--Carlo simulations there is no clear
transition from regular oscillations to chaotic ones or from stable
expansion to oscillations, as observed in gas and Fokker--Planck
models. However, the present results are consistent with results
obtained by Takahashi \& Inagaki (1991) for stochastic Fokker--Planck
model (stochastic binary formation and energy generation), by Makino
(1996) for $N$--body simulations and by Giersz
\& Spurzem (1997) for anisotropic gaseous model with fully 
self-consistent Monte-Carlo treatment of binary population. 

The Monte--Carlo code is 
at least $10^5$ times faster than the $N$--body one for $N=32768$ with 
special--purpose hardware (Makino 1996). Thus it becomes possible to 
run several different models to improve statistical quality of the data 
and run individual models with $N$ larger then $100000$. The 
Monte--Carlo scheme can be regarded as a method which lies in the middle 
between direct $N$--body and Fokker--Planck models and combines most 
advantages of both methods.

\bigskip
This work was supported in part by the Polish National 
Committee for Scientific Research under grant 2--P304--009-06.

\bigskip

\section{Anisotropic Gaseous Models of Star Clusters by R. Spurzem}

\bigskip

Gaseous sphere models of star clusters have been powerful tools to
examine the dynamical evolution of star clusters. The physical
nature of gravothermal
collapse (Lynden-Bell \& Eggleton 1980) and the existence of gravothermal
oscillations (Bettwieser \& Sugimoto 1984) were detected using
gaseous models. They use a phenomenological heat flux equation in
order to close the moment equations of the Boltzmann equation with
a Fokker-Planck local collisional term. Anisotropy is taken into
account for the second order moments (radial and tangential
velocity dispersions, compare Spurzem 1994). It was shown
that in pre- and post-collapse for particle numbers ranging
from $N=250$ to $N=10000$ there is a very good agreement between
results of gaseous models, direct numerical solutions of the
orbit-averaged Fokker-Planck equation and direct $N$-body simulations
(cf Giersz \& Spurzem 1994, Spurzem \& Aarseth 1996). Similarly
comparisons between gaseous and other models of star clusters with
two different masses provided results in fair agreement with each
other (Spurzem \& Takahashi 1995). 

The previously cited work, however, only examined isolated systems
consisting of point masses. Surprisingly, and in contrast to other models,
no results have been  published recently including a tidal boundary
and stellar evolution effects using gaseous models. In the course
of an ongoing project for doing that, we provide here the first 
multi-mass gaseous models including a tidal boundary, for comparison with
results of the other methods (Heggie, this paper, Sect. 1). 
To include a tidal boundary is less
straightforward than in Fokker-Planck or $N$-body models.
In the $N=10000$ direct $N$-body simulation of
Spurzem \& Aarseth (1996) most escapers suffered their last scattering
well inside the core. Guided by that we employ the ansatz that at
each radius $r$ of a spherical star cluster the mass loss rate by
escape across the tidal boundary can be determined by
$${\delta\rho\over\delta t} = {k x \rho\over t_{\rm esc}}. $$
Here $\rho$ is the local stellar mass density, $t_{\rm esc}=
r/v_{\rm esc}$ the local timescale for escape with the escape
speed $v_{\rm esc}=\sqrt{\Phi_t - \Phi}$, using the potential
difference between $r$ and the tidal radius $r_t$. Assuming
a Schwarzschild-Boltzmann distribution $x$ is the fraction of stars
located in the escape region of velocity space (for the determination
of $x$ it is possible to distinguish between an energy and apocentre
criterion for escape as in the anisotropic Fokker-Planck models of 
Takahashi (this paper, Sect. 2),
and the actual computation of $x$ is done by approximating a
three-dimensional volume integral over an ellipsoidal figure by
integrations over a certain combination of
volumes with rectangular boundaries). Finally $k$ is the
``filling degree'' of the escaper region, taking into account that
most stars in that region quickly escape. So far we tested
single mass systems with $10^4$ solar-mass stars and the
here presented multi-mass models with $6\cdot 10^4$ solar masses
in total. Best agreement
was reached by choosing $k=1\cdot 10^{-7} M/{\rm M}_\odot$,
where $M$ is the total mass of the system. $k$ should be determined
by a diffusion equation; this and details of the previously described
procedure will be given elsewhere. 

Another important improvement of gaseous models is inspired by the
revival of H\'enon type Monte-Carlo models (Giersz 1997). Star formation
is believed to start with a high fraction of so-called primordial
binaries,
most of which are destroyed in three- and four-body encounters
during the dynamical evolution. Such a high amount of binaries cannot
be treated by the Fokker-Planck equation anymore. Previous work either
was not self-consistent (Hut et al. 1992), or assumed in the single mass
model some very preliminary cross sections for close binary-binary encounters
(Gao et al. 1991). Direct $N$-body simulations with many binaries suffer
from the high computational cost (Aarseth \& Heggie 1992). A new
self-consistent model including a stochastic binary component treated
by H\'enon's Monte-Carlo method, including relaxation (i.e. dynamical
friction with the single stars and binary-binary relaxation) and
close encounters was proposed and successfully tested (Spurzem \& Giersz
1996). It will provide a self-consistent
quasi $N$-body representation for large
$N$-body systems with a huge number of binaries for extremely small
computational cost. Like in a direct $N$-body simulation
three- and four-body encounters can be integrated in regularized
coordinates and effects of stellar evolution and collisions could be
incorporated in the model, which is subject of future work.

\bigskip
This work was supported in part by DFG (German Science Foundation) grant
Sp 345/3-3, 10-1.

\end{document}